\documentclass{IET-Conf-Paper}

\makeatletter
\def\input@path{{./tables/}}
\makeatother

\graphicspath{{figures/}}

\usepackage{lipsum,booktabs,amsmath,amssymb,subfig,bbm}

\newcommand{\CCV}{\mathrm{CCV}}
\newcommand{\CCA}{\mathrm{CCA}}
\newcommand{\etaA}{\eta_{\mathrm{A}}}
\newcommand{\etaV}{\eta_{\mathrm{V}}}

\newcommand{\U}{\mathcal{U}}

\newcommand{\lb}{\left (}
\newcommand{\rb}{\right )}
\newcommand{\Ufix}{\mathcal{U}^{\mathrm{Fix}}}

\newcommand{\Sopt}{\mathcal{S}^{\mathrm{Opt.}}}
\newcommand{\Iopt}{\mathcal{I}^{\mathrm{Opt.}}}

\newcommand{\vh}{\vspace{0.5em}}

\newcommand{\Ptot}{P_{\mathrm{Ttl.}}}
\newcommand{\Ineut}{I_{\mathrm{Ntrl.}}}
\newcommand{\Tclarke}{T_{\mathrm{Clarke}}}

\newcommand{\abg}{\alpha \beta \gamma }

\begin{document}

This paper is a preprint of a paper submitted to the Proceedings of Power Electronics, Machines and Drives 2024 (PEMD 2024) and is subject to Institution of Engineering and Technology Copyright. If accepted, the copy of record will be available at IET Digital Library.

\newpage

\title{RECONFIGURABLE POWER CONVERTERS WITH INCREASED UTILIZATION FOR UNBALANCED POWER DISTRIBUTION SYSTEM APPLICATIONS}

\author{Matthew Deakin\corr, Xu Deng}

\address{{School of Engineering, Newcastle University, Newcastle-upon-Tyne, UK}\\ \email{matthew.deakin@newcastle.ac.uk}}


\begin{abstract}
A low-cost reconfiguration stage connected at the output of balanced three-phase, multi-terminal ac/dc/ac converters can increase the feasible set of power injections substantially, increasing converter utilization and therefore achieving a lower system cost. However, the approach has yet to be explored for phase unbalance mitigation in power distribution networks, an important application for future energy systems. This study addresses this by considering power converter reconfiguration's potential for increasing the feasible set of power transfers of four-wire power converters. Reconfigurable topologies are compared against both conventional four-wire designs and an idealised, fully reconfigurable converter. Results show that conventional converters need up to 75.3\% greater capacity to yield a capability chart of equivalent size to an idealised reconfigurable converter. The number and capacity of legs impact the capability chart's size, as do constraints on dc-side power injections. The proposed approach shows significant promise for maximizing the utilization of power electronics used to mitigate impacts of phase unbalance.
\end{abstract}

\maketitle

\section{Introduction}

The electrification of demand and uptake of distributed generation in domestic properties will lead to a significant increase in load on low voltage power (LV) distribution networks \cite{speakman2022low}. In many regions, domestic customers are connected to a single phase supply, or have a three-phase supply feeding single phase loads \cite{ma2020review}. In the absence of active phase balancing, it can be expected the vast majority of assets that would have to be replaced due to thermal constraints would be under-utilized prior to their replacement. This is because it is the phase with the highest loading, rather than the average loading, which limits operation of the distribution system components such as substation transformers \cite{ma2016quantification}.

\begin{figure}
\centering
\includegraphics[width=0.48\textwidth]{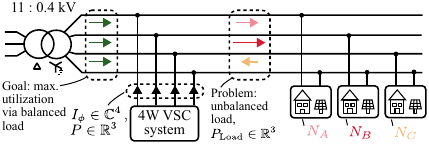}
\caption{A four-wire (4W) voltage source converter (VSC) system can be used to balance phase currents by injecting unbalanced active powers. Balanced phase currents ensures that network assets can be fully utilized.}
\label{f:system_setup_pemd}
\end{figure}

As a result, there has been rapidly growing interest in approaches to increase utilization of congested branches prior to their reinforcement. Proposed approaches include static reallocation of single-phase loads \cite{pereira2021phase}, dynamic customer phase reallocation via static transfer switch-based `phase shifting devices' \cite{shahnia2014voltage,liu2021load,cui2023two}, the control of single-phase customer generators \cite{girigoudar2022integration}, or the use of distributed energy resources solutions interfaced through a three-phase four-wire (4W) power converter. The latter approach is shown in Fig.~\ref{f:system_setup_pemd}, with works published considering such an approach including standalone Static Compensator (STATCOM)-based approaches \cite{de2016modeling}, battery energy storage \cite{stecca2022battery}, or back-to-back `soft open points' (SOPs) \cite{cui2023two}. However, the approach is not limited to these and could also include any other dc resource--e.g., an LVDC cable, a three-phase electric vehicle fast charger, or industrial motor drive. These four wire (4W) voltage source converter (VSC)-based systems can support increased utilization by monitoring phase currents at a congested branch (e.g., a substation transformer), then dynamically injecting unbalanced power to reduce loading on heavily loaded phases.

If the phase unbalance mitigation provided by a 4W VSC system changes in time, then the loading on the individual legs of the interfacing ac/dc converter will also vary. As a result, a 4W VSC system appears to be an attractive proposition for the use of the multiplexing approach described in \cite{deakin2023optimal,deakin2023multiplexing}. In particular, it is proposed that 4W VSC systems are constructed of an arbitrary number of legs, whose current-carrying capacity is not necessarily uniform, and whose output can be connected to any of the four wires of the power distribution system. By allowing the capacity connected to each of the four wires to change dynamically, the feasible set of power transfers can be augmented. The purpose of this work is to explore this feasible set (i.e., the capability chart), and explore how different sizing strategies can increase the size of this set. 

There are a few works that have explored the use of power converter reconfiguration for mitigating phase unbalance. In \cite{lou2020new}, the authors describe a `phase changing SOP', assumed to be constructed using a multi-terminal `phase shifting device' in recent works \cite{cui2023two}. However, we show in this work that a three-leg, three-phase inverter with equally sized legs cannot benefit from reconfiguration on the output. Alternatively, works on balanced operation of SOPs and their capability charts have been considered in detail in \cite{deakin2023multiplexing,deakin2023optimal}. However, to the authors' knowledge, there are no works that consider capability charts of reconfigurable power converters for arbitrary unbalance mitigation, including the critical effect of neutral currents. This is an important and timely gap, as the proposed reconfiguration approach promising to lower the cost of power electronics by increasing utilization substantially \cite{deakin2023multiplexing}, and thereby providing a route for network operators to install more cost-effective active solutions to address distribution system congestion.

The major contribution of this work is to address this gap by proposing and calculating capability charts for the proposed four-wire reconfigurable power converters. This includes numerical approaches to determine the size of these capability charts, and corresponding visualizations of these sets. The aim is provide metrics that can demonstrate the improved flexibility of the reconfiguration approach.

This work is structured as follows. In Section~\ref{s:methodology}, we describe the proposed topology and define corresponding capability charts, then describe how the area (or volume) of these charts can be calculated numerically to demonstrate performance improvement for a given 4W VSC design. In Section~\ref{s:results}, a range of case studies are presented to highlight the improvements that can be expected as a result of the proposed reconfiguration. Finally, salient conclusions are drawn in Section~\ref{s:conclusion}.

\section{Methodology}\label{s:methodology}

\begin{figure}\centering
\subfloat[Standalone 4W VSC System]{\includegraphics[height=4.6em]{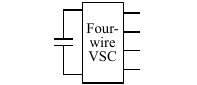}\label{f:vsc_standalone}}~
\subfloat[Interconnected 4W VSC System]{\includegraphics[height=4.6em]{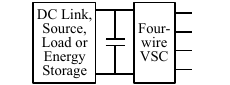}\label{f:vsc_interconnected}}
~
\caption{In this work, we classify Four Wire VSC Systems as either (a) Standalone, which cannot transfer power to or from the dc subsystem (but can transfer power between phases), or (b) Interconnected, with some resource connected to the dc subsystem to enable net power transfer across the VSC.}
\label{f:vsc_varieties}
\end{figure}

In this work, two basic structures are considered for the 4W VSC system, as shown in Fig.~\ref{f:vsc_varieties}. A Standalone 4W VSC, as shown in Fig.~\ref{f:vsc_standalone}, consists of only a four-wire VSC and dc capacitor, and so the power balance restricts the currents that can be injected (i.e., the sum of active powers injected into the network must have value zero). The standalone device is similar to a STATCOM, which has been used for providing dynamic reactive power compensation in power systems for several decades, and which can transfer power between phases to mitigate high currents. In contrast, the Interconnected 4W VSC systems, as shown in Fig.~\ref{f:vsc_interconnected}, also has a dc-side component connected to enable a non-zero power injection from the VSC into the ac network.

In this Section, we outline how the capability chart of a both of these classes of 4W VSC systems can be increased by adding the proposed reconfiguration output stage. Firstly, in Section~\ref{ss:mechanism} the mechanism by which this reconfiguration increases the capability chart is revisited, highlighting with a simple example how the proposed approach can be beneficial under unbalanced operation. Subsequently, in Section~\ref{ss:capability}, capability charts are defined as a set, with it highlighted how the size of these sets change as a function of total VSC capacity. Finally, in Section~\ref{ss:numerical}, numerical routines are described for calculating the size of these sets.

\subsection{Proposed Four-Wire Reconfigurable Power Converter}\label{ss:mechanism}

\begin{figure}\centering
\subfloat[Conventional design]{\includegraphics[width=0.84\linewidth]{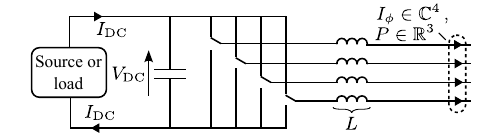}\label{ff:power_poles_nominal}}
~\\
\subfloat[Proposed design with reconfiguation and leg resizing]{\includegraphics[width=0.84\linewidth]{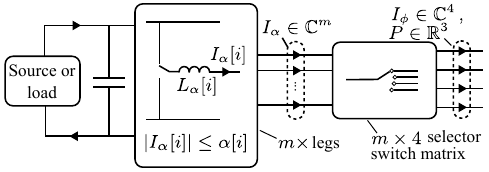}\label{ff:power_poles}}
\caption{Conventional and proposed four-wire VSC system designs. The conventional design (a) has four legs hard-wired to the ac output wires, where the proposed reconfigurable design (b) has any number of legs $m$ with arbitrary current ratings $ \alpha $ whose output can be reconfigured to redistributed current carrying capacity as necessary.}
\label{f:p_poles}
\end{figure}

The essence of the proposed approach has been described in detail in \cite{deakin2023multiplexing} for a balanced, multi-terminal back-to-back VSC system. In a conventional VSC system (Fig.~\ref{ff:power_poles_nominal}), the output of each of the equally-sized half-bridge (HB) legs are hard-wired to the four output wires; the maximum current that can be injected into any one of the phases is static. In contrast, the proposed approach can have any number of legs, $ m $, whose pu size $ \alpha $ can be varied (at the design stage), and which can be connected to any of the outputs to vary the HB capacity connected to a given phase on-the-fly. It is worth noting that if the mission profile for the 4W VSC does not vary with time, then the proposed approach will not provide a benefit.

There are two major differences between the balanced and unbalanced cases. Firstly, neutral currents can be neglected in the balanced case. In contrast, as we demonstrate in Section~\ref{ss:standalone}, neutral currents are often large and so must be calculated for an accurate determination of the capability chart. Secondly, individual legs of the VSC are considered, rather than three-phase VSC units as in, e.g., \cite{deakin2023multiplexing}.

For the purposes of this work, but without loss of generality, we consider the main purpose of such a system is to inject only active power $ P $ (and not reactive power), for the following reasons. Firstly, analysis of smart meter data suggests that modern domestic customers have very good power factor--for example, it is suggested in \cite{vanin2023analysis} that a power factor of unity ought to be assumed for simulation tasks. If this is the case, then injecting reactive power will increase (rather than decrease) loading on each phase, and so is not useful for addressing thermal congestion. Secondly, so long as the converter has sufficient capacity, issues with reactive power can be addressed with single phase assets injecting reactive power (where active power phase rebalancing is not possible without changing the load point). Finally, reactive power can be mitigated dynamically by other low-cost means (e.g., switched capacitor banks). Nevertheless: reactive power could be also considered in future applications, considering applications such as voltage unbalance mitigation.

\subsubsection{Example Reconfigurable Converter}

\begin{figure}
\centering
\includegraphics[width=0.47\textwidth]{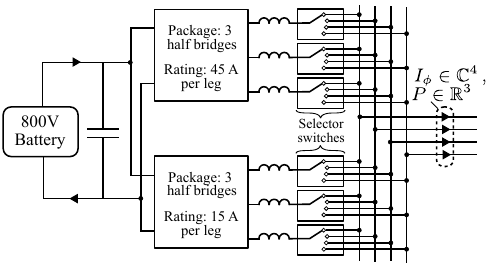}
\caption{An example reconfigurable 4W VSC system with battery energy storage system and two HB packages.}
\label{f:power_poles_xmpl}
\end{figure}

An example of a reconfigurable converter, constructed of two HB triplet packages of 45~A and 15~A respectively are shown in Fig.~\ref{f:power_poles_xmpl}. Such a converter can be connected in various operating modes.
\begin{itemize}
\item The converter can operate under balanced operation with maximum power output. In this case, 60A is connected to phase 1--3; the neutral has no capacity allocated to it. In this mode, the converter is acting like a conventional three-wire, three-phase converter.
\item The converter could be used to inject 90~A of active power into any of one of phases 1--3 to mitigate a temporary overload, with 90 A connected to the neutral (for the return current). In this mode, the converter is acting like a single-phase converter. 50\% extra current can be injected into the phase which the HBs are connected to as compared to the balanced converter.
\item The battery is disconnected for maintenance. 45~A of capacity is connected to two of phases 1--3; 30~A to the other phase; and 60~A to the neutral to enable phase rebalance currents (as shown in Section~\ref{ss:standalone}, the neutral current is typically larger than the phase currents). The converter is acting as a four wire standalone VSC system; depending on the optimal operation, the 45~A and 30~A converters can be reallocated.
\end{itemize}
In summary, the selector switches can enable a wide range of unbalanced operation, with a higher utilization of the HBs as compared to a static design. In the next Section, we propose the use of capability charts to systematically study the full range of operation (as compared to the specific operating points highlighted in this example).

\subsection{Defining Capability Charts}\label{ss:capability}

A capability chart can be defined as the feasible operating region of a given power converter, i.e., the feasible currents and/or voltages under which it can operate. This has been defined for balanced reconfigurable converters in \cite{deakin2023optimal,deakin2023multiplexing}. However, unbalanced current injections by a converter necessitates consideration of neutral currents, and so in this section we revisit capability chart for the four-wire configuration.

For a 4W power converter operating with a fixed output voltage and no dc-side constraints (e.g., assuming a large dc link capacitor) the power converter's operating range is determined by the current rating of the capacity connected to each wire $ I_{\phi}^{\max} \in \mathbb{R}_{+}^{4} $ (in Amps or pu). The phasors representing the phase currents in each leg of the converter $ I_{\phi} \in \mathbb{C}^{4} $ must therefore satisfy
\begin{equation}\label{e:i_constraint}
|I_{\phi}[i]| \leq I_{\phi}^{\max} [i] \; \forall \; i \in \{ 1,\,2,\,3,\,4 \}\,.
\end{equation}
For such a system, consider the $ m $ per-unit leg capacities $ \alpha \in \mathbb{R}^{m} $, such that 
\begin{equation}\label{e:alpha_sum}
\sum_{i=1}^{m} \alpha [i] = 1\,.
\end{equation}
A variable binary matrix $ B \in \{0,1\}^{4\times m} $ has $ m $ columns that each represent the state of the $ m $ one-to-four selector switches, with the sum of each column of $ B $ having value 1 (to ensure each leg is connected to just one output wire). Then, the capacity connected to each output wire $ I_{\phi}^{\max} $ is simply the sum of the capacities of the legs connected to that wire, i.e.,
\begin{equation}\label{e:max_current}
I_{\phi}^{\max} = I_{\mathrm{base}} B \alpha \,,
\end{equation}
where the current $ I_{\mathrm{base}} $ has value of the sum of the capacity of all legs (in Amps). Note that, if a converter is not reconfigurable, then $ B $ is replaced with an identity matrix.

For a four-wire converter injecting in-phase currents (i.e., active powers) into a stiff positive sequence voltage with fixed phase voltage magnitudes $ |V_{0}| $, the per-phase powers $ P\in \mathbb{R}^{3} $ are linked to those phase currents $ I_{\phi} $ according to the linear relation
\begin{subequations}\label{e:p2i}
\begin{align}
I_{\phi} (P) [i]  &= \dfrac{P[i]}{|V_{0}|a^{(1-i)}} \,, \quad i \in \{1,\,2,\,3\}\,,\\
I_{\phi}(P) [4]  &= - \sum _{i=1}^{3 } I_{\phi}(P)[i]\,.
\end{align}
\end{subequations}
where $ a =e^{2\pi \jmath/3} $ is the phase rotation operator. Note that the neutral current $ \Ineut $ flows through the fourth wire in \eqref{e:p2i},
\begin{equation}\label{e:i_neutral_defn}
\Ineut = I_{\phi}[4]\,,
\end{equation}
and that the neutral is at the reference voltage at 0~V. As a result, the neutral cannot inject either active or reactive power (which is why there are only three components in the powers $ P $). 

The capability chart can therefore defined as the feasible set of power injections $ P $ that satisfy the power converter operating constraints, i.e.,
\begin{equation}\label{e:capability_chart}
C = \{ P: \eqref{e:i_constraint},\,\eqref{e:max_current},\, \eqref{e:p2i} \}\,.
\end{equation}
The advantage of the capability chart in terms of active powers $ P $ (instead of current phasors $ I_{\phi} $) is that, in the general case of arbitrary $ P $ (i.e., an Interconnected 4W VSC), the dimension of the set ($\mathrm{dim}(C) = 3$) is the same as the number of variables in $ P $. 

The total power injected into the ac grid, $ \Ptot $, is
\begin{equation}\label{e:p_tot}
\Ptot = P[1] + P[2] + P[3]\,.
\end{equation}
For the case of a Standalone 4W VSC (Fig.~\ref{f:vsc_standalone}), ac-side powers must balance, i.e., $\Ptot$ must have value of zero; for an Interconnected 4W VSC, $ \Ptot $ can have a non-zero value.

Given the definition of a capability chart as a set, it is natural to consider the size of the set as a property of interest when comparing converters with different $ \alpha $. The capability chart volume (CCV) or capability chart area (CCA) are
\begin{equation}\label{e:ccv_cca_defn}
\CCV = \int _{C} dV\,,\qquad \CCA = \int _{C} dA\,,
\end{equation}
with the former used for an Interconnected 4W VSC, and the latter a Standalone 4W VSC (whose capability chart $ C $ is two-dimensional, as the three powers $P$ are constrained to lie in the plane $\sum P[i]=0$). 

It is worth noting that the CCV and CCA scale with the cube and square of the converter power rating, respectively (as can be observed by considering the units of the integrals). For a given converter, the scaling factors $ \etaV,\, \etaA $ that yield the same volume or area, respectively, for given CCA or CCVs are \cite{deakin2023multiplexing}
\begin{equation}\label{e:equiv_sizes}
\qquad \etaV = \sqrt[3]{ \dfrac{\CCV_{2}}{\CCV_{1}} }\,, \etaA = \sqrt{ \dfrac{\CCA_{2}}{\CCA_{1}} } \,.
\end{equation}
For example, if the value of the CCA of a second converter is double that of a first, then the first converter would need to be a factor of $ \etaA=\sqrt{2} $ larger to have the same CCA.

\subsubsection{HB Sizing Case Studies}\label{sss:benchmarks}
To explore potential benefits of the proposed approach for 4W current unbalance mitigation, we consider four sizing and reconfiguration approaches.
\begin{itemize}
\item An $ m $-leg reconfigurable converter, each with uniform leg sizes of capacity $ 1/m $~pu, denoted $ \U (m) $.
\item Three-leg and four-leg converters with uniform HB leg capacities and legs hard-wired to phase wires, denoted $ \Ufix(3),\,\Ufix(4) $ respectively. Note that $ \Ufix(3) $ does not have a neutral for return current.
\item Designs with leg capacities chosen to maximise the Standalone system CCA, $ \Sopt_{4} $; or the Interconnected system CCV, $ \Iopt_{4} $. The sizes of each leg was chosen by evaluating the value of $ \alpha $ that maximises the CCA or CCV using a meshgrid of all $ \alpha $ at a resolution of 0.01 pu. CCA and CCV values for each $ \alpha $ are determined using the method of Section~\ref{sss:brute_method}.
\item An idealised converter, denoted $ \Omega $--i.e., a design for which leg capacity can be allocated continuously, providing an upper bound on CCA and CCV values \cite{deakin2023multiplexing}.
\end{itemize}
The capability chart for the idealised design $ \Omega $ is found by replacing the leg reconfiguration constraint \eqref{e:max_current} with the continuous constraint
\begin{equation}\label{e:max_current_idealised}
|I_{\phi} (P) [i]| \leq 1\,.
\end{equation}

\subsection{Calculating Capability Chart Area or Volumes}\label{ss:numerical}

Determining the CCA or CCV requires the evaluation of the multi-dimensional integrals \eqref{e:ccv_cca_defn}. In general, such integrals can be computationally challenging \cite{deakin2023multiplexing}. However, considering uniform converters $ \U(m) $ or converters with four legs, these integrals can be calculated very conveniently as we demonstrate.

\subsubsection{Brute Force Grid Methods}\label{sss:brute_method}

In this work, CCA or CCV values are determined by first considering a regular grid of points over a region $ R $ that is known to completely enclose the capability chart $ C $. An indicator function $ F(P,\,\alpha) $ is then defined that returns value 1 if the point lies within the capability chart and zero otherwise. The CCA or CCV is then conveniently found via a normalised sum of the indicator function values at all points across the grid.

If there are four legs of the converter, then the indicator function is
\begin{equation}\label{e:indicator_four}
F_{4}(P,\alpha ) = 
\begin{cases}
1 & \mathrm{if}\quad \alpha[i] \leq |I_{\phi}[i]| \; \forall \; i \in \{1,\,2,\,3,\,4\}\,,\\
0 & \mathrm{otherwise\,,}
\end{cases}
\end{equation}
where $ P $ is linked to phase currents \eqref{e:p2i}, $ I_{\phi} $ has been assumed as in per-unit, and (with slight abuse of notation) $ \alpha $ and $ |I_{\phi}| $ have been ordered according to their size. For converters with uniform leg sizing $ \U (m)$, the indicator function is
\begin{equation}\label{e:indicator_uniform}
F_{\U}(P,\alpha ) = 
\begin{cases}
1 & \mathrm{if}\quad \sum _{i=1}^{4} \mathrm{ceil}( m|I_{\phi}(P)[i]| ) \leq m \,,\\
0 & \mathrm{otherwise\,,}
\end{cases}
\end{equation}
where $ \mathrm{ceil}(\cdot) $ is the `ceiling' function (i.e., non-integer arguments are rounded up to the nearest integer).

\subsubsection{Evaluating the Boundary of the Capability Chart}\label{sss:boundary_method}
A complementary approach for exploring the capability chart $ C $ is by finding its boundary. This has two potential uses for analysis and visualization. Firstly, it can be used as a method to determine the CCA or CCV. For example, if the radius $ r $ of the boundary of the capability chart $ C $ of a Standalone 4W VSC is known as a function of the angle $ \theta $ in the plane, then
\begin{equation}\label{e:radius_integrate}
\CCA = \int_{\theta=0}^{2\pi} r(\theta ) d \theta\,.
\end{equation}
This could be an advantage as it reduces the dimension of the grid of points that must be evaluated over (e.g., from 2 to 1 for CCAs). Secondly, when the total power injection $ \Ptot $ is non-zero, the boundary is not necessarily connected to the interior of the capability chart (as we show in Section~\ref{ss:interconnected}). It therefore provides different information than that which is provided by considering the size of the set (i.e., the CCA or CCV) alone.

To efficiently determine the radius at the boundary $r$, it is proposed to consider the capability charts in $ \abg $ co-ordinates via the Clarke transformation,
\begin{equation}\label{e:trn_clarke}
\hat{P} = \Tclarke P\,,
\end{equation}
where the Clarke transformation matrix $ \Tclarke $ is
\begin{equation}\label{e:t_clarke_mat}
\Tclarke = \sqrt{\dfrac{2}{3}}\begin{bmatrix}
1 & -1/2 & -1/2 \\
0 & \sqrt{3}/2 & -\sqrt{3}/2 \\
1/\sqrt{2} & 1/\sqrt{2} & 1/\sqrt{2} \\
\end{bmatrix}\,.
\end{equation}
The transformed powers $ \hat{P} $ in $ \abg $ co-ordinates can be interpreted as follows. The third element, $ \hat{P}[3] $, is proportional to the total power injected $ \Ptot $, as
\begin{equation}\label{e:ptot2z}
\hat{P}[3] = \dfrac{\Ptot}{\sqrt{3}}\,,
\end{equation}
as can be seen by considering \eqref{e:p_tot}, \eqref{e:t_clarke_mat}, \eqref{e:trn_spherical}. This means that co-ordinates $ \hat{P}[1],\,\hat{P}[2] $ lie in the plane of the Standalone VSC system's capability chart and correspond to unbalanced power injections. Secondly, the radius $ r $ in polar or cylindrical co-ordinates,
\begin{equation}\label{e:r_defn}
r^{2} = \hat{P}[1]^{2} + \hat{P}[2]^{2}\,,
\end{equation}
is proportional to magnitude of the neutral current $ |\Ineut |$. This can be seen by considering the magnitude of the neutral current \eqref{e:p2i} mapped through the Clarke transformation \eqref{e:trn_clarke}. From \eqref{e:r_defn}, constraints on neutral currents are therefore circles in $ \abg $ co-ordinates (rather than ellipses in nominal co-ordinates). Therefore, visualizations in $ \abg $ co-ordinates have additional symmetry and explainability.

For an Interconnected VSC, it is useful to represent powers $ \hat{P} $ in spherical co-ordinates,
\begin{subequations}\label{e:trn_spherical}
\begin{align}
\hat{P}[1] &= r_{\mathrm{Sph.}} \sin (\theta ) \cos (\psi ) \\
\hat{P}[2] &= r_{\mathrm{Sph.}} \sin (\theta ) \sin (\psi ) \\
\hat{P}[3] &= r_{\mathrm{Sph.}} \cos (\theta ) \,,
\end{align}
\end{subequations}
where $ r_{\mathrm{Sph.}} $ is the spherical radius, $ \psi $ the azimuth angle, and $ \theta $ the polar angle. In contrast to cylindrical co-ordinates, if a point at $ r_{\mathrm{Sph.}} $ is feasible for a given $ \psi,\,\theta $, then all points between the origin and $ r_{\mathrm{Sph.}} $ will also be feasible.

The radius at the boundary of the capability chart can be found either by defining an appropriate indicator function and then using a root-finding method, or via an optimization to maximise $ r $ or $ r_{\mathrm{Sph.}} $. For example, for a fixed $ \psi $ and $ \theta $, the radius $ r_{\mathrm{Sph.}} $ at the boundary of the capability chart can be determined as the solution of a mixed-integer linear program,
\begin{subequations}\label{e:max_r}
\begin{align}
\max & \; r_{\mathrm{Sph.}} \\
\mathrm{s.t.}\;  \eqref{e:i_constraint},\, \eqref{e:max_current},\, & \eqref{e:p2i},\, \eqref{e:trn_clarke},\, \eqref{e:trn_spherical}\,.
\end{align}
\end{subequations}

\subsubsection{Capability Chart Discontinuities}\label{sss:discontinuities}

If the current injected into one of the wires has value of zero, then no leg capacity needs to be connected to that wire. This means that at there are regions of the capability chart which, according to the CCA and CCV definitions, do not contribute to the size of the set, as these subsets will be of lower dimension. Nevertheless, as we show in Section~\ref{s:results}, these points are non-trivial, and could be very useful for supporting some mission profiles.

The subspaces of these points in a given co-ordinate system can be found by setting one of the powers $P$ to zero (or the neutral current $\Ineut$), then solving for the parameters that capture that locus. For example, for a fixed azimuth angle $ \psi $ (in spherical co-ordinates \eqref{e:trn_spherical}), the powers are a function of polar angle $ \theta $ as
\begin{equation}\label{e:delta_angle}
\Tclarke^{\intercal} r \begin{bmatrix}
\cos (\psi) & 0 \\
\sin (\psi) & 0 \\
0 & 1
\end{bmatrix}
\begin{bmatrix}
\sin(\theta )\\
\cos(\theta )
\end{bmatrix}
= P\,.
\end{equation}
The angle $ \theta $ that will yield the $ i $th element of $ P $ to be zero can be found by setting the $ i $th row of \eqref{e:delta_angle} to be zero, then solving for $ \theta $.

\section{Results}\label{s:results}

In this section, we evaluate the size of the capability charts for the converter sizing approaches given in Table~\ref{t:alpha_designs}, considering both Standalone and Interconnected 4W VSC operation. This demonstrates how the proposed approach can improve converter flexibility. Both qualitative and quantitative properties of these capability charts are considered to explore the performance of these designs.

\begin{table}
\centering

\caption{{The number of legs $ m $ and leg capacities $ \alpha $ for\\each design.}}

\begin{tabular}{lll}

\toprule

Symbol & $ m $ (no. legs) & $ \alpha $ (leg capacities) \\

\midrule

$ \Sopt_{4} $ & $ 4 $ & (0.12, 0.22, 0.26, 0.4) \vh \\
$ \Iopt_{4} $ & $ 4 $ & (0.13, 0.21, 0.3, 0.36) \vh \\
$ \U(m) $ & $ m $ & $\lb\lb \frac{1}{m} \rb^{m\times} \rb $ \vh \\
$\U^{\mathrm{Fix}}(3)$ & $ 3 $ & $\lb \lb \frac{1}{3} \rb^{3\times} \rb $ \vh \\
$\U^{\mathrm{Fix}}(4)$ & $ 4 $ & $\lb \lb \frac{1}{4} \rb^{4\times} \rb $ \vh \\
$ \Omega $ & n.a. & $ \lim_{m \to \infty} \U(m) $ \\

\bottomrule

\end{tabular}
\label{t:alpha_designs}
\end{table}

\subsection{Capability Chart Areas for Standalone Four-Wire VSCs}\label{ss:standalone}

\begin{figure}
\centering
\includegraphics[width=0.44\textwidth]{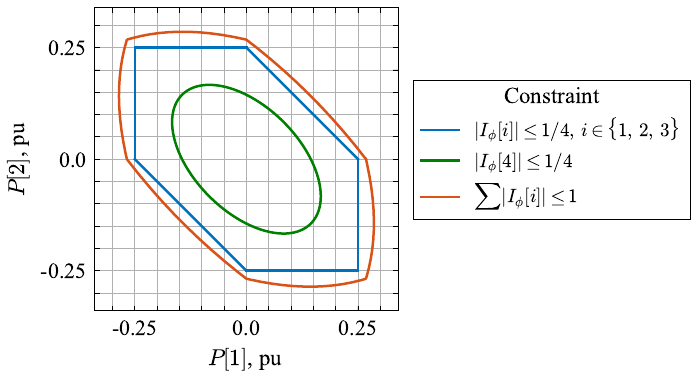}
\caption{Constraints that create capability chart $ C $ for a fixed converter $ \Ufix(4) $ and the idealised converter $ \Omega $.}
\label{f:pltBalancedConstraintsGeom}
\end{figure}

In the first instance, we consider the boundary of the capability chart $ C $ for a Standalone VSC for a conventional, fixed design $ \Ufix(4) $ and compare this against the idealised design $ \Omega $, as shown in Fig.~\ref{f:pltBalancedConstraintsGeom}. The constraints on power injection on the phase legs form a hexagon (in direct analogy to the balanced three-terminal SOP case \cite{deakin2023optimal}). In contrast, the constraint on the neutral current forms an ellipse. Interestingly, this ellipse is completely contained within the hexagon. This highlights that a converter with all four legs sized symmetrically highly underutilizes the three phase legs--only the neutral leg is active on the boundary of the capability chart. The idealised capability chart completely encapsulates the 0.25~pu hexagon, and so even if the neutral current were not the limiting factor (or an alternative means was provided for the neutral return, such as a split-capacitor), the idealised design $ \Omega $ would still improve performance as compared to a fixed four-leg VSC $ \Ufix(4) $.

\begin{figure}\centering
\subfloat[Nominal co-ordinates]{\includegraphics[height=12.5em]{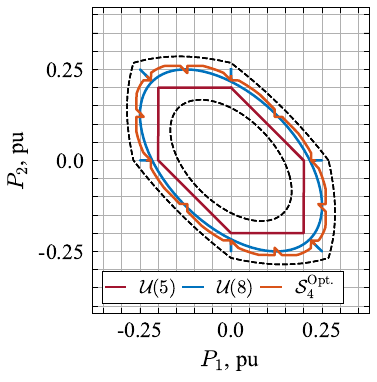}\label{f:pltGallerySnap_xyz}}
\subfloat[$ \abg $ co-ordinates]{\includegraphics[height=12.5em]{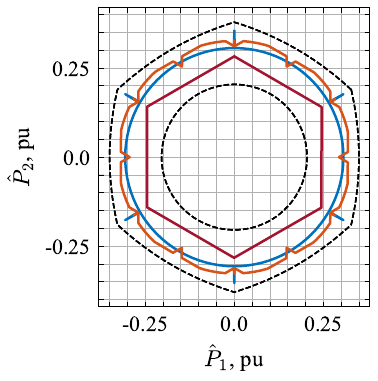}\label{f:pltGallerySnap_abg}}
\caption{Capability charts for a 5- and 8- converter uniform designs, $ \U(5),\,\U(8) $ and the optimal 4-leg reconfigurable design $ \Sopt_{4} $. For comparison, the feasible set of the idealised and fixed four-leg design are also plotted as dashed lines.}
\label{f:pltGallerySnap}
\end{figure}

In Fig.~\ref{f:pltGallerySnap}, the boundary of the capability charts for three designs are illustrated in nominal and $\abg$ co-ordinates, with the capability chart of the benchmark cases $ \Omega,\,\Ufix(4) $ shown in dashed lines. It can be seen in this figure that the capability chart varies significantly depending on the sizing $ \alpha $. For the 5-converter uniform design $ \U(5) $, 0.4~pu is allocated to the neutral and so the capability chart is a hexagon. For a uniform, 8-converter uniform design $ \U(8) $, the capability chart is enlarged significantly, although, it can be seen that that design is mostly contained within the optimal 4-converter design $ \Sopt_{4} $. Additionally, Fig.~\ref{f:pltGallerySnap} shows that many of the properties of capability chart areas of reconfigurable balanced VSCs can be observed: non-convexity; existence of capability charts as strict supersets of other capability charts; and, large increases in area compared to a conventional design \cite{deakin2023optimal}.

\begin{figure}
\centering
\includegraphics[width=0.47\textwidth]{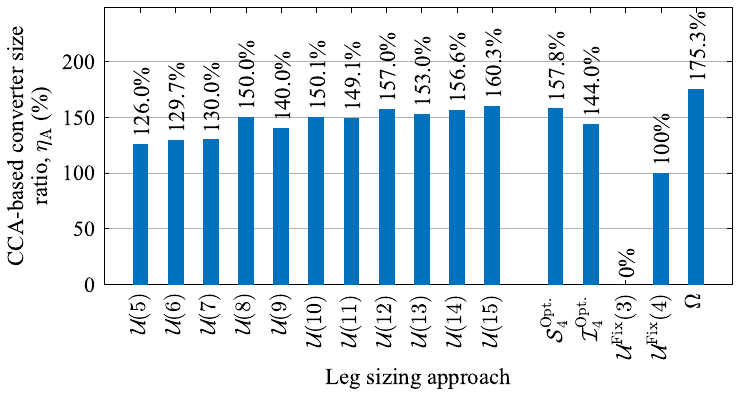}
\caption{Bar chart of the CCA-based converter size ratio $ \etaA $ (from \eqref{e:equiv_sizes}) for a range of leg sizing approaches.}
\label{f:plt_PR_table_ccas_pemd}
\end{figure}

The CCA-based converter sizing ratio $\etaA$ for the leg sizing approaches collected in Table~\ref{t:alpha_designs} are plotted in Fig.~\ref{f:plt_PR_table_ccas_pemd}, considering uniform designs with up to 15 legs $ m $. From this figure, the poor performance of conventional designs $ \Ufix(3),\,\Ufix(4) $ can clearly be seen--for the former, there is no neutral return, and so power can only be transferred between two phases at a time (and so the CCA is trivially zero). For the latter, it can be seen that the fixed converter $ \Ufix(4) $ requires 57.8\% capacity increase to achieve the same capacity as the optimal design $ \Sopt_{4} $, with an upper limit of an increase 75.3\% as compared to the idealised converter $ \Omega $. It is also interesting to see that designs with uniformly-sized converters are relatively inefficient at increasing the capacity of the CCA, with 15 uniformly-sized converters required to achieve a larger CCA than the optimal design $ \Sopt(4) $. This highlights how effective choice of leg sizes $ \alpha $ can yield substantially increased flexibility, echoing previous findings \cite{deakin2023optimal,deakin2023multiplexing}.

\subsection{Capability Chart Volumes for Interconnected Four-Wire VSCs}\label{ss:interconnected}

\begin{figure}\centering
\subfloat[Nominal co-ordinates]{\includegraphics[height=10em]{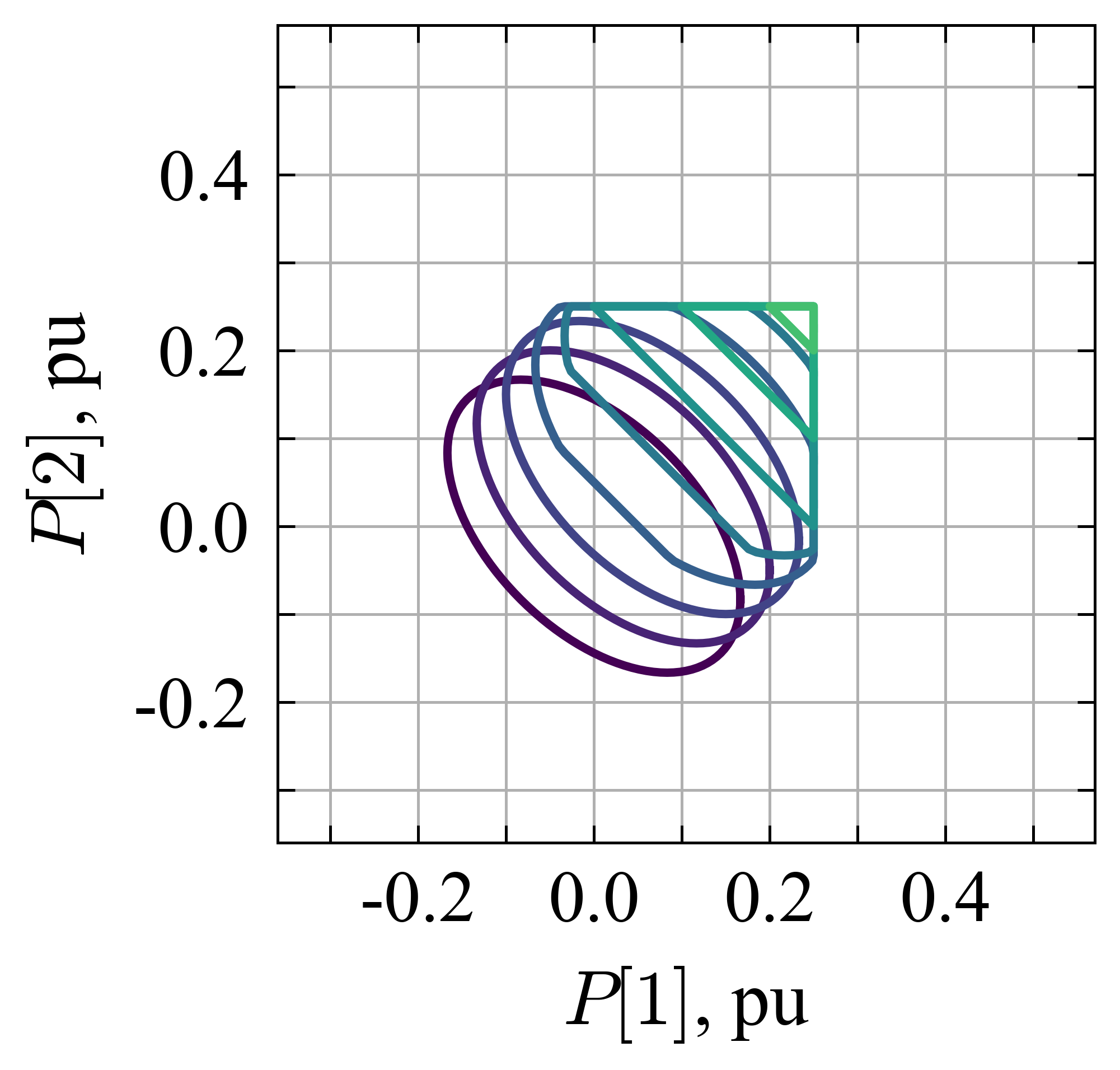}\label{f:pltFixed4Lcc_P_fix25}}
~
\subfloat[$\abg$ co-ordinates]{\includegraphics[height=10em]{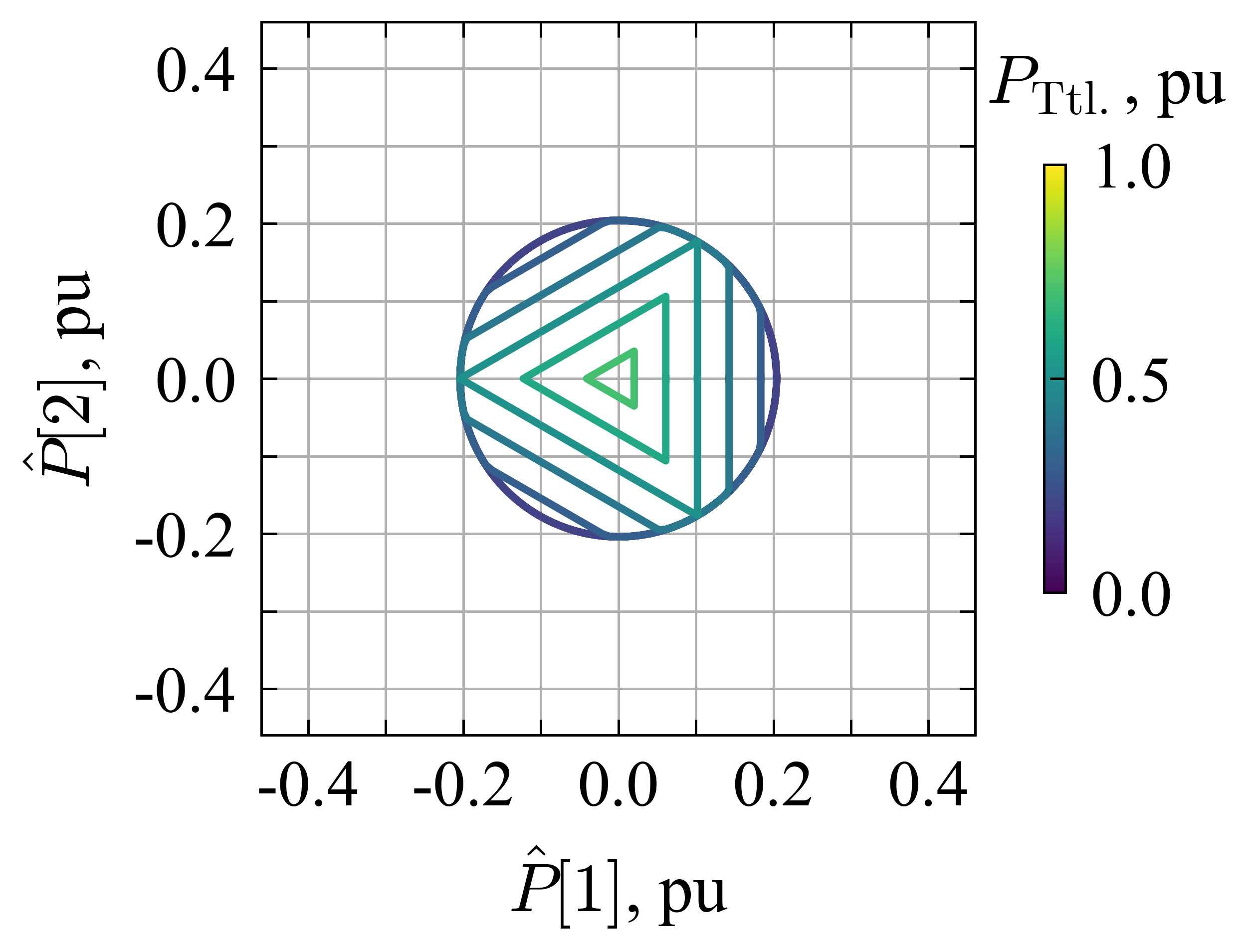}\label{f:pltFixed4Lcc_Ph_fix25}}
\caption{Boundary of the capability chart $ C $ for fixed values of total power injection $ \Ptot $ for the conventional design $ \Ufix(4) $.}
\label{f:pltFixed4Lcc}
\end{figure}

\begin{figure}\centering
\subfloat[Nominal co-ordinates ]{\includegraphics[height=10em]{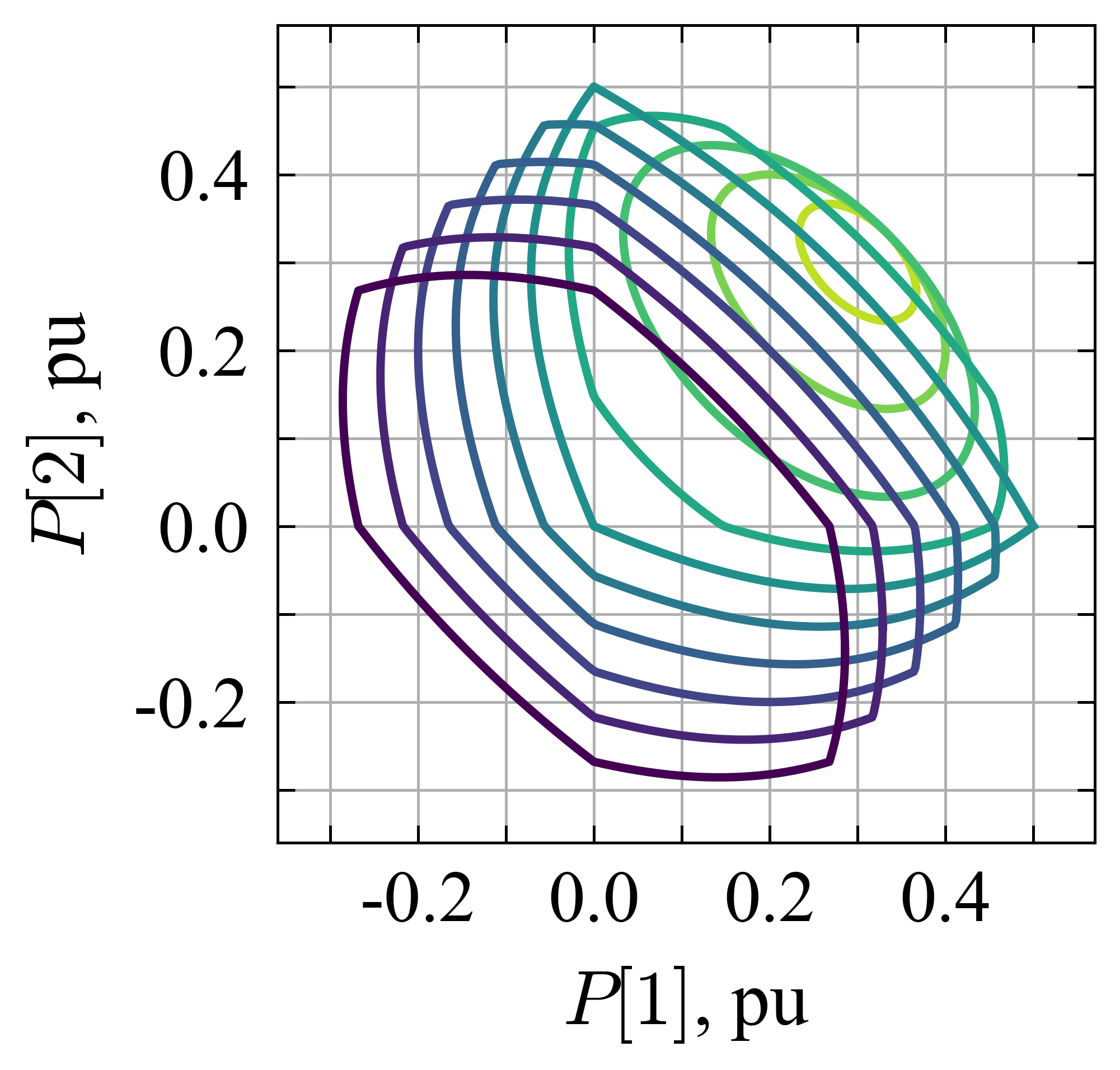}\label{f:plt4lVscChartAnlytc}}
~
\subfloat[$\abg$ co-ordinates]{\includegraphics[height=10em]{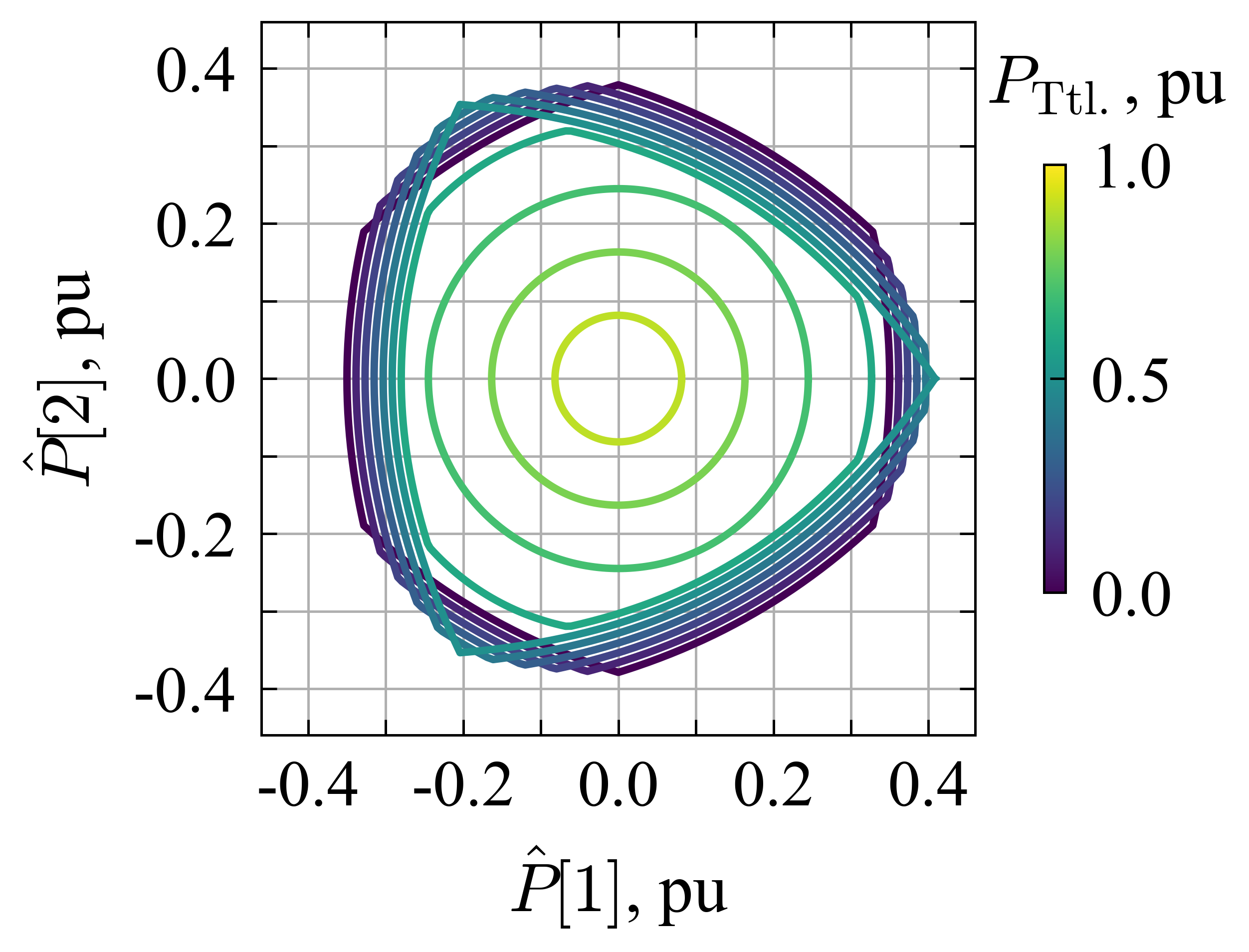}\label{f:plt4lVscChartAnlytc_h}}
\caption{Boundary of the capability chart $ C $ for fixed values of total power injection $ \Ptot $ for the idealised design $ \Omega $.}
\label{f:plt4lVscChartAnlytc_pair}
\end{figure}

The boundary of the capability chart for the conventional, fixed design $ \Ufix(4) $ and the idealised design $ \Omega $ are plotted in Fig.~\ref{f:pltFixed4Lcc} and Fig.~\ref{f:plt4lVscChartAnlytc_pair}, respectively, for a range of values of total power injection $\Ptot$. As expected, it can be observed a total power injection $ \Ptot $ with value zero matches the Standalone 4W VSC (Fig.~\ref{f:pltBalancedConstraintsGeom}); but, as the total power injection $ \Ptot $ increases, the shape of the capability chart changes significantly. For example, as $ \Ptot $ increases above 0.5 pu, the set of feasible injections quickly reduces, reaching a single point at 1~pu injection. A power injection $ \Ptot $ of 0.5~pu leads to the greatest per-phase power injection potential of 0.5~pu (by comparison, if the power injection $\Ptot$ is 1~pu, the per-phase power injection is 1/3~pu, i.e., 50\% lower).

\begin{figure}\centering
\subfloat[$ \psi=0 ^{\circ}$]{\includegraphics[height=9.4em]{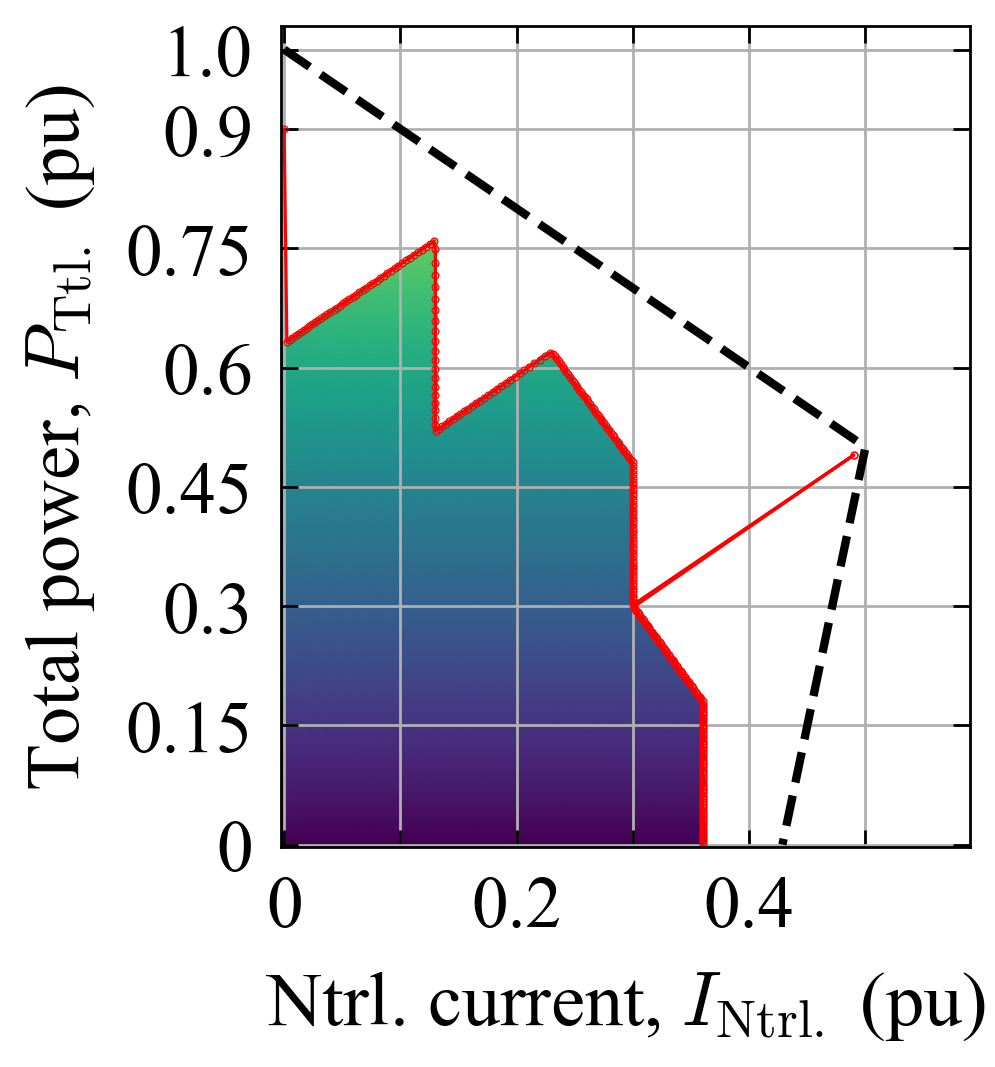}\label{f:plt4L4Pmc_spherical_a}}
\subfloat[$ \psi=3 ^{\circ}$]{\includegraphics[height=9.4em]{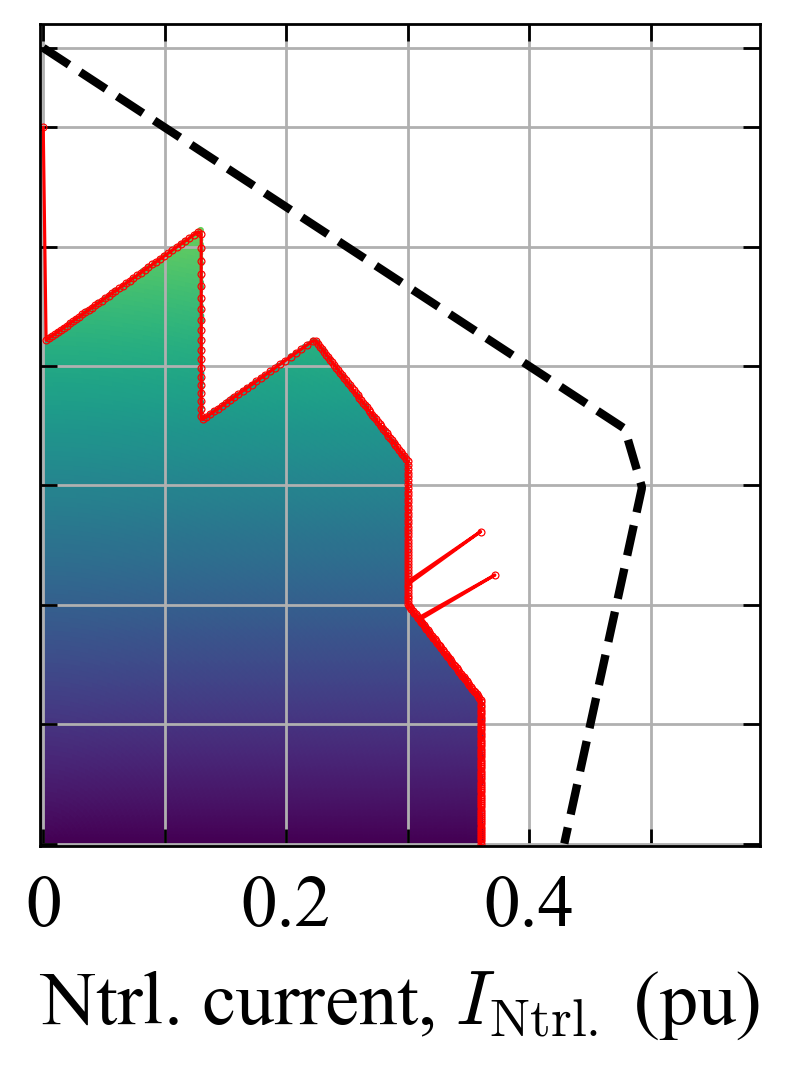}\label{f:plt4L4Pmc_spherical_b}}
\subfloat[$ \psi=45 ^{\circ}$]{\includegraphics[height=9.4em]{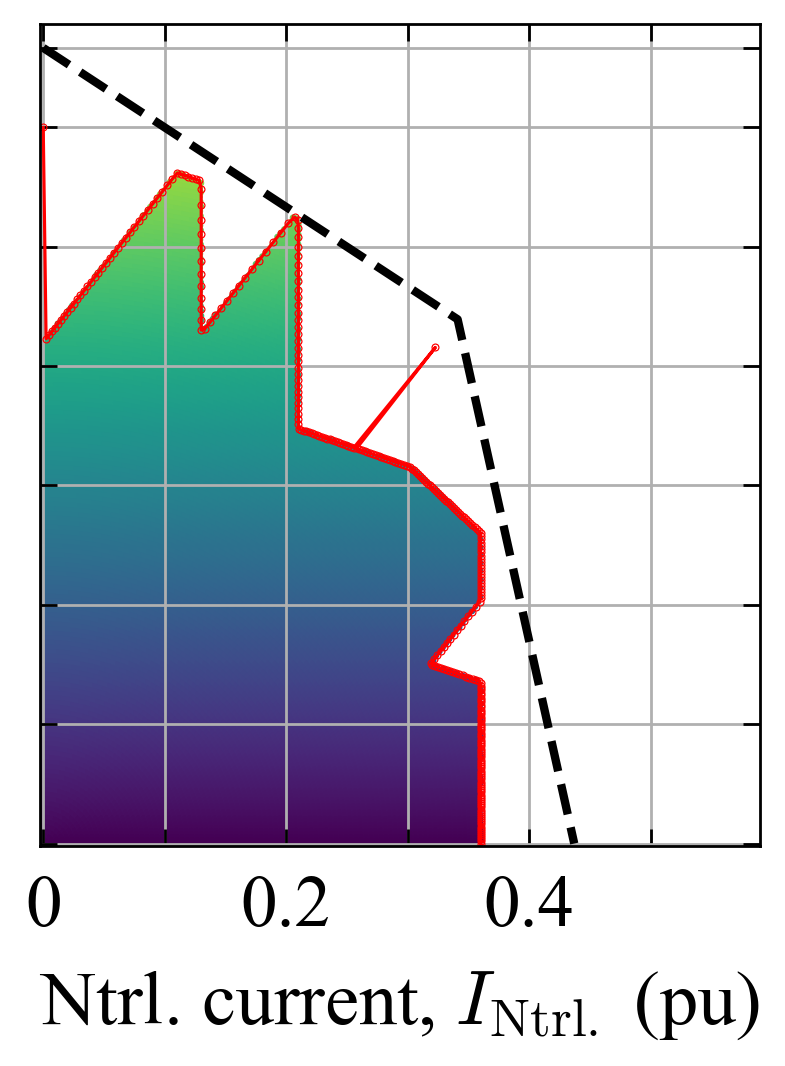}\label{f:plt4L4Pmc_spherical_c}}
\caption{Capability charts $ C $ for three values of azimuth angle $ \psi $ for the interconnected VSC optimal four-leg design $ \Iopt_{4} $.}
\label{f:plt4L4Pmc_spherical}
\end{figure}

For reconfigurable designs, the capability charts of Interconnected 4W VSC systems are more complex than those of Standalone systems. For example, the capability chart for three angles of azimuth $ \psi $ are plotted in Fig.~\ref{f:plt4L4Pmc_spherical} (only the first quadrant is plotted due to symmetry). The filled areas are calculated using Section~\ref{sss:brute_method}'s brute-force method, with the red outer boundary line calculated by determining the boundary in spherical co-ordinates (Section~\ref{sss:boundary_method}); the black dashed line indicating the boundary of the idealised design. It can be observed that the Interconnected VSC can have isolated points, as described in Section~\ref{sss:discontinuities}. Due to the complex non-convexities in the capability chart, only snapshots at given $ \psi $ are given (as compared to the full capability charts which can be plotted in Fig.~\ref{f:pltFixed4Lcc} and Fig.~\ref{f:plt4lVscChartAnlytc_pair} for the conventional and idealised designs, respectively).

\begin{figure}\centering
\subfloat[$ \Ptot=0.45 $ pu]{\includegraphics[height=9.4em]{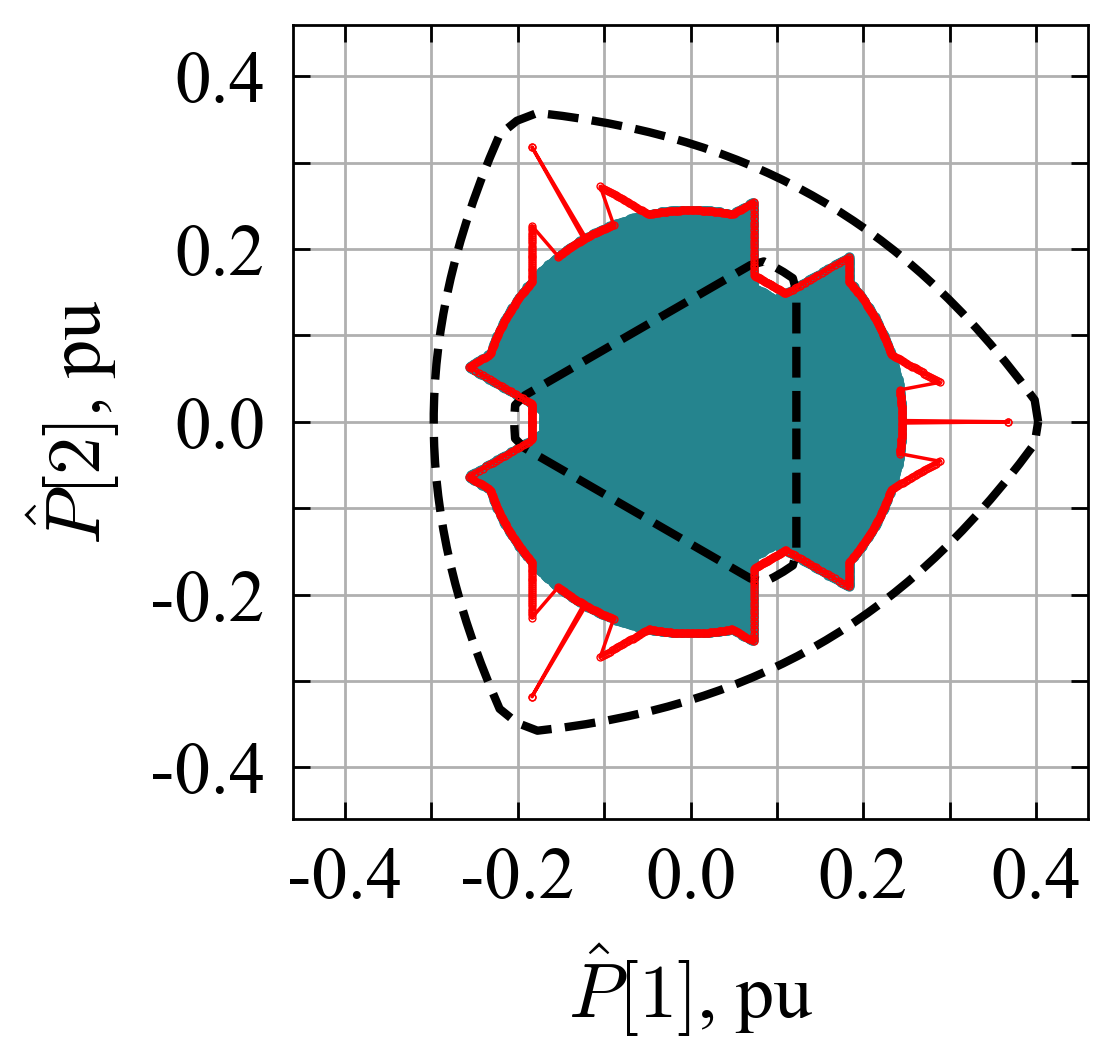}\label{f:plt4L4Pmc_a}}
\subfloat[$ \Ptot=0.6 $ pu]{\includegraphics[height=9.4em]{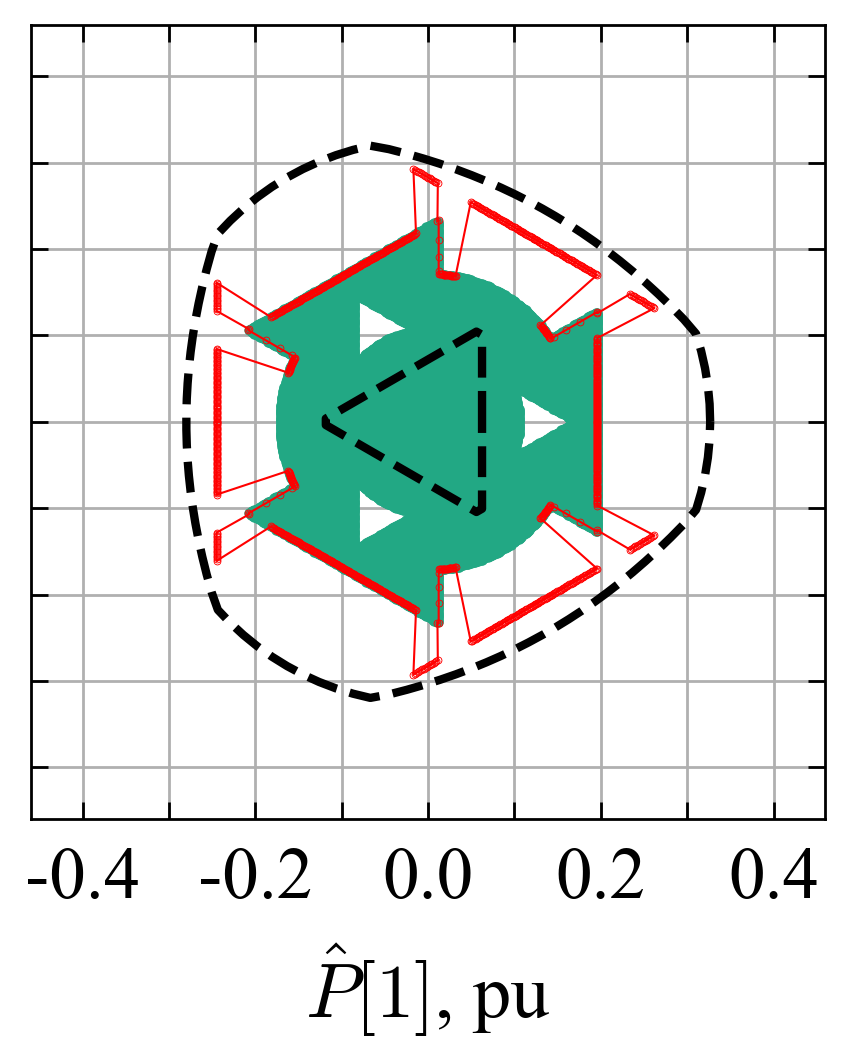}\label{f:plt4L4Pmc_b}}
\subfloat[$ \Ptot=0.75 $ pu]{\includegraphics[height=9.4em]{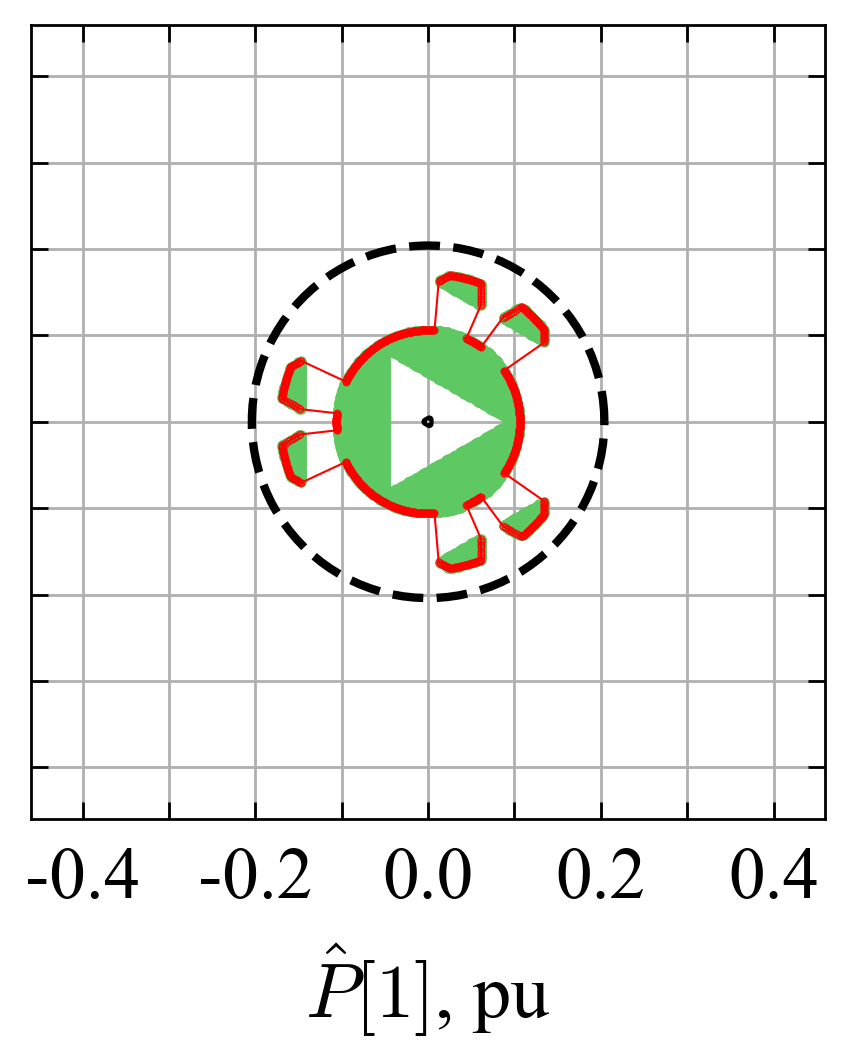}\label{f:plt4L4Pmc_c}}
\caption{Capability charts $ C $ for three values of $ \Ptot $ for the interconnected VSC optimal four-leg design $ \Iopt_{4} $.}
\label{f:plt4L4Pmc}
\end{figure}

For a fixed value of $ \Ptot $, the capability charts can have more complexity again. The capability charts for three values of $ \Ptot $ are shown in Fig.~\ref{f:plt4L4Pmc}, with solid coloured areas are plotted using Section~\ref{sss:brute_method}'s brute-force methods (coloured area); red lines calculated using Section~\ref{sss:boundary_method}'s boundary method in cylindrical co-ordinates; and, black dashed lines showing the boundary of the capability chart of the idealised design $ \Omega $ and benchmark $ \U^{\mathrm{Fix}}(4) $. There are two main differences between these spherical and cylindrical capability charts representations. Most strikingly, there can be `holes' in the capability charts when plotted for a fixed $ \Ptot $, as shown in Fig.~\ref{f:plt4L4Pmc_b} and Fig.~\ref{f:plt4L4Pmc_c}. For example, close to the origin, it may be the case that even with the smallest leg connected to the neutral, there is not sufficient capacity to inject $ \Ptot $ in an (almost) balanced way; so, there must be a reasonable amount of unbalanced power injected to enable the full value of $ \Ptot $. 

Secondly, it is interesting to note that when there is a fixed $ \Ptot $, there can be isolated points (Fig.~\ref{f:plt4L4Pmc_a}), isolated lines  (Fig.~\ref{f:plt4L4Pmc_b}), and isolated areas (Fig.~\ref{f:plt4L4Pmc_c}). This could lead to more complex operation as compared to a conventional power converter--a more complicated procedure would need to be run to determine the connection of the converters as the mission profile changes, for example.

\begin{figure}
\centering
\includegraphics[width=0.4\textwidth]{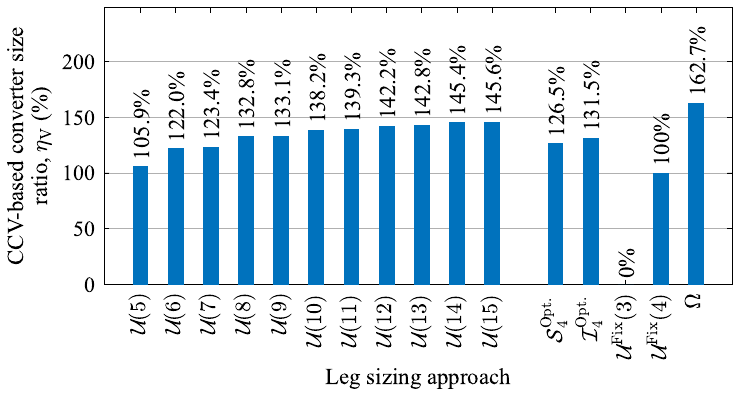}
\caption{Bar chart of the CCV-based converter size ratio $ \etaV $ (from \eqref{e:equiv_sizes}) for a range of leg sizing approaches.}
\label{f:plt_AFE_table_ccvs_pemd}
\end{figure}

Finally, the CCV-based converter size ratios $ \etaV $ for various designs are shown in Fig.~\ref{f:plt_AFE_table_ccvs_pemd}. Compared to the Standalone 4W VSC case, the gap between the size ratio $ \etaV $ of the conventional design $ \Ufix(4) $ and the idealised design $ \Omega $, the potential for improvement is slightly reduced to 62.7\%. Nevertheless, it remains a significant increase, and this ratio is still larger than that of the three-terminal SOP considered in \cite{deakin2023optimal} (which has a value of $ \etaA $ of 150\% for the idealised design). In contrast to the Standalone VSC, uniform sizing is also much more effective at increasing the CCV compared to optimal design $ \Iopt_{4} $, with the nine-converter design $ \U(9) $ surpassing the CCV of the optimal design $ \Iopt_{4} $.

\section{Conclusion}\label{s:conclusion}

The reconfiguration of power converters using low-cost selector switches in applications with variable mission profiles has previously shown significant promise in balanced distribution systems. In this work, the approach has been shown to be effective as a potential approach for current unbalance mitigation, with a conventional design requiring between 62.7\% and 75.3\% increased capacity to match an idealised converter. Efficient numerical methods have been given to determine the feasible sets and boundary of the capability charts, with more complex capability charts as compared to previously considered reconfigurable converters, particularly when the VSCs are interconnected to a dc-side power source or load. 

It is concluded that power converter reconfiguration can be an effective tool for reallocating capacity of individual legs of power converters in several different applications. Future work could explore hardware approaches to implement the proposed reconfiguration approach to consider more clearly the practical challenges of implementing the selector switches. In addition, other optimal design and operational approaches will be considered to further increase converter flexibility and utilization.

\section{Acknowledgements}

M. Deakin was supported by the Royal Academy of Engineering under the Research Fellowship programme. X. Deng was supported by Newcastle University Academic Track (NUAcT) Fellowship  scheme. 

\section{References}

%\bibliography{master_bib_pemd}
%\bibliographystyle{IEEEtran}

%\bibliography{master_bib_pemd}
%\bibliographystyle{IEEEtran}
% Generated by IEEEtran.bst, version: 1.14 (2015/08/26)

\end{document}